\documentclass[12pt,times]{article}
\usepackage{epsfig}
\usepackage{amsfonts,amssymb}
\usepackage{lscape}
\usepackage{fullpage}
\usepackage{natbib}
\usepackage{algorithm, algorithmic}
\bibpunct{[}{]}{;}{a}{,}{,}

\newcommand{\mbf}[1]{\mathbf{#1}}
\newcommand{\be}{\begin{equation}}
\newcommand{\ee}{\end{equation}}

\newcommand{\comment}[1]{}

\newcommand{\myfigref}[1]{Fig.~\ref{#1}}

\begin{document}

\title{A Spectral Graph Approach to Discovering Genetic Ancestry}
\author{Ann B. Lee, Diana Luca, and Kathryn Roeder}
\maketitle

\section*{Abstract}

Mapping human genetic variation is fundamentally interesting in fields
such as anthropology and forensic inference.  At the same time
patterns of genetic diversity confound efforts to determine the
genetic basis of complex disease.  Due to technological advances it is
now possible to measure hundreds of thousands of genetic variants per
individual across the genome.  Principal component analysis (PCA) is
routinely used to summarize the genetic similarity between subjects.
The eigenvectors are interpreted as dimensions of ancestry. We build
on this idea using a spectral graph approach.  In the process we draw
on connections between multidimensional scaling and spectral kernel methods.  
Our approach, based on a spectral embedding derived from the
normalized Laplacian of a graph, can produce more meaningful
delineation of ancestry than by using PCA.  
 The method is stable to
outliers and can more easily incorporate different similarity measures
of genetic data than PCA.  We illustrate 
a new algorithm for genetic clustering and association analysis on a large, genetically heterogeneous sample.

\section*{Introduction}

Human genetic diversity is of interest in a broad range of contexts,
ranging from understanding the genetic basis of disease, to
applications in forensic science. Mapping clusters and clines in the
pattern of genetic diversity provides the key to uncovering the
demographic history of our ancestors.  
To determine the genetic basis of complex disease, individuals are measured at large numbers of genetic variants across the genome as part of
the effort to discover the variants that increase liability to complex
diseases such as autism and diabetes.  

Genetic variants, called alleles, occur in pairs, one inherited from
each parent.  High throughput genotyping platforms routinely yield genotypes for
hundreds of thousands of variants per sample.  These are usually
single nucleotide variants (SNPs), which have two possible alleles,
hence the genotype for a particular variant can be coded based on
allele counts (0,1 or 2) at each variant.  The objective is to
identify SNPs that either increase the chance of disease, or are
physically nearby an SNP that affects disease status.  

\begin{figure}
\includegraphics[angle=0,width=7in]{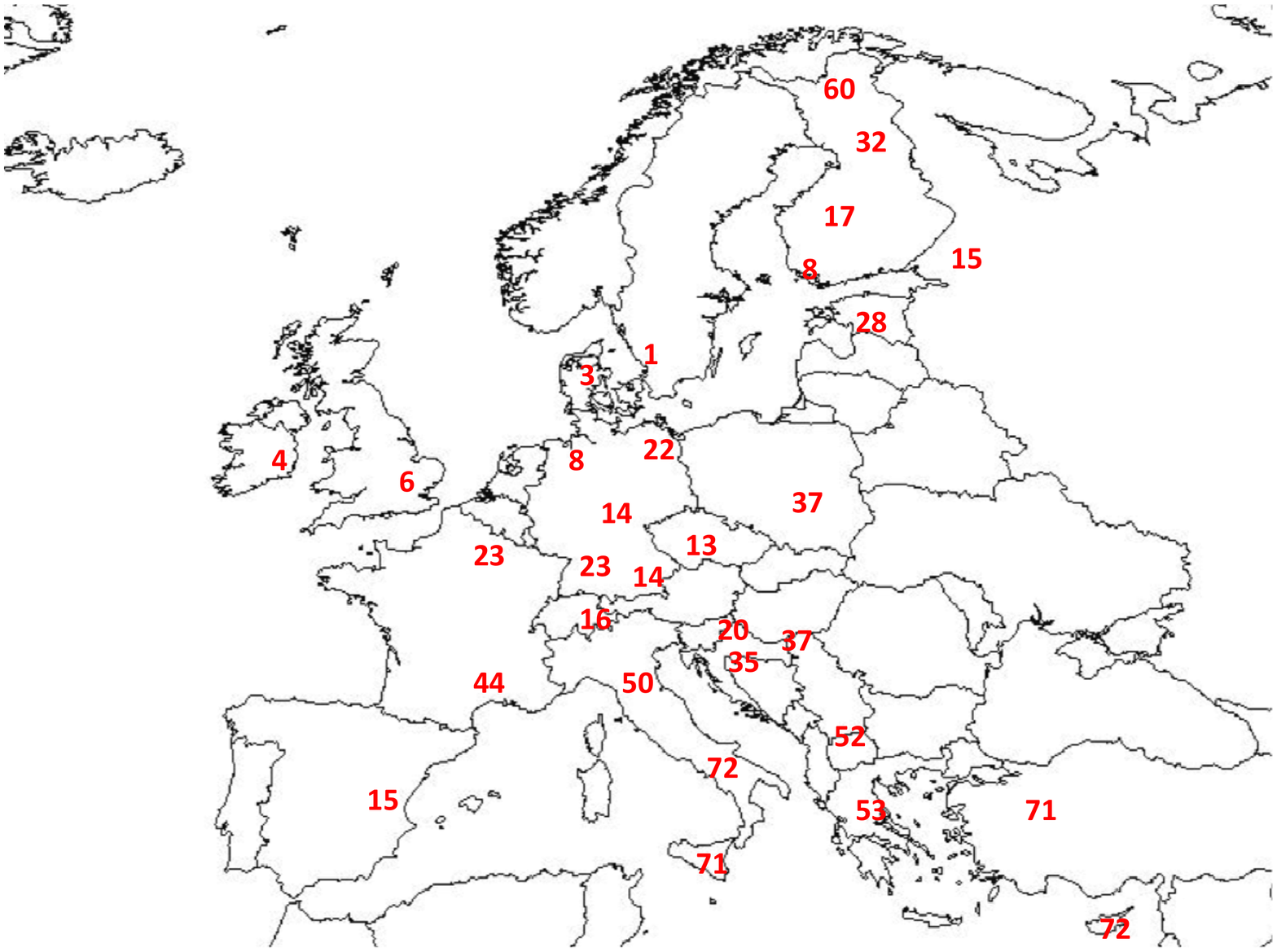}
\caption{Percent of adult population who are lactose intolerant
(http://www.medbio.info/Horn/Time). A gradient runs from north to south, 
correlating with the spread of the lactase mutation. Finland provides 
an exception to the gradient due to the Asian influence in the north.}
\label{fig:lactase}
\end{figure}

Due to demographic, biological, and random forces, variants differ in
allele frequency in populations around the world \citep
{cavalli}. An allele that is common in one geographical or
ethnic group may be rare in another.  For instance, the O blood type
is very common among the indigenous populations of Central and South
America, while the B blood type is most common in Eastern Europe and
Central Asia \citep{cavalli}.  
The lactase mutation, which facilitates
the digestion of milk in adults, occurs with much higher frequency in
northwestern Europe than in southeastern Europe (\myfigref{fig:lactase}). Ignoring
the structure in populations leads to spurious associations in
case-control genetic association studies due to differential
prevalence of disease by ancestry.

Although most SNPs do not vary dramatically in allele frequency across
populations, genetic ancestry can be estimated based on allele counts
derived from individuals measured at a large number of SNPs.  An
approach known as structured association clusters individuals to
discrete subpopulations based on allele frequencies \citep{pritchard}. 
This approach suffers from two limitations: results are
highly dependent on the number of clusters; and realistic populations
do not naturally resolve into discrete clusters.  If fractional
membership in more than one cluster is allowed the calculations
becomes computationally intractable for the large data sets currently
available.  A simple and appealing alternative is principal component
analysis (PCA \citep{cavalli,price,patterson}), or principal component
maps (PC maps).  This approach summarizes the genetic similarity
between subjects at a large numbers of SNPs using the dominant
eigenvectors of a data-based similarity matrix. Using this
``spectral'' embedding of the data a small number of eigenvectors is
usually sufficient to describe the key variation. The PCA framework
provides a formal test for the presence of population structure based
on the Tracy-Widom distribution \citep{patterson, johnstone}. Based on
this theory a test for the number of significant eigenvectors is
obtained. 

In Europe, eigenvectors displayed in two dimensions often reflect the
geographical distribution of populations \citep{heath,novembre}.
There are some remarkable examples in the population genetics
literature of how PC maps can reveal hidden structures in human
genetic data that correlate with tolerance of lactose across Europe
\citep{tishkoff}, migration patterns and the spread of farming
technology from Near East to Europe \citep{cavalli}. Although these
stunning patterns can lead to overinterpretation \citep{novembre2},
they are remarkably consistent across the literature.

In theory, if the sample consists of $k$ distinct subpopulations,
$k-1$ axes should be sufficient to differentiate these
subpopulations. In practice, finding a dimension reduction that
delineates samples collected worldwide is challenging.  For instance,
analysis of the four core HapMap samples [African, Chinese, European,
and Japanese; HapMap-Consortium, 2005] using the classical principal
component map \citep{patterson} does not reveal substructure within
the Asian sample; however, an eigenmap constructed using only the
Asian samples discovers substructure \citep{patterson}.  Another
feature of PCA is its sensitivity to outliers \citep{t1d}.  Due to
outliers, numerous dimensions of ancestry appear to model a
statistically significant amount of variation in the data, but in
actuality they function to separate a single observation from the bulk
of the data.  This feature can be viewed as a drawback of the PCA
method.

Software is available for estimating the significant eigenvectors via
PCA (Eigenstrat \citep{price}, smartpca \citep{patterson}, or
GEM~\citep{t1d}).  For population-based genetic association studies,
such as case-control studies, the confounding effect of genetic
ancestry can be controlled for by regressing out the eigenvectors
\citep{price,patterson}, matching individuals with similar genetic
ancestry \citep{t1d,rosenbaum}, or clustering groups of individuals
with similar ancestry and using the Cochran-Mantel-Haenszel test.  In
each situation, spurious associations are controlled better if the
ancestry is successfully modeled.


To overcome some of the challenges encountered in constructing a
successful eigenmap of the genetic ancestry, we propose a spectral
graph approach.  These methods are more flexible than PCA (which
 can be considered as a special case) and allow
for different ways of modeling structure and similarities in data. The basic idea is 
to represent the population as a weighted graph, where the vertex set is 
comprised by the subjects in the study, 
and the weights reflect the degree of similarity
 between pairs of subjects. 
The graph is then embedded in a lower-dimensional space using the top eigenvectors 
of a function of the weight matrix. Our approach utilizes a spectral embedding derived from
the so-called normalized graph Laplacian.  Laplacian eigenmaps and spectral graph methods 
are well-known and widely used in machine learning but unfamiliar to many classically trained statisticians and biologists. The goals of this work are to:
\begin{itemize}
\item demonstrate the use of spectral graph methods in the analysis of population structure in genetic data
\item emphasize the connection between PCA methods used in population genetics and more general spectral methods used in machine learning
\item develop a practical algorithm and version of Laplacian eigenmaps for genetic association studies
\end{itemize}
We proceed by discussing
the link between PCA, multidimensional scaling (MDS) and
spectral graph methods.  We then present a practical scheme for 
determining the number of significant dimensions of ancestry 
by studying the gap statistic of the eigenvalues of the graph Laplacian. 
We conclude with a presentation of the new
algorithm, which is illustrated via analyses of the POPRES data
\citep{nelson} and simulated data with spurious associations.

\section*{Methods}
 
\subsection*{Spectral embeddings revisited. Connection to MDS and
  kernel PCA.}
We begin by making the connection between multidimensional scaling
(MDS) and the principal component (PC) method explicit: Suppose $Z$ is
an $n \times p$ data matrix, with rows indexed by $n$ subjects and
columns indexed by $p$ biallelic SNP markers. Center each column
(marker) to have mean $0$; denote the centered data matrix $X=AZ$
where $A=I-\frac{1}{n}\mbf{1}\mbf{1}^t$ is an $n \times n$ centering
matrix. The elements of the $i$th row of $X$ represent the genetic
information for subject $i$, $\mbf{x}_i=(x_{i1},\ldots,x_{ip})$.

A singular value decomposition of $X$ gives $$X=U \Gamma V^t,$$ 
where $\Gamma$ is a diagonal matrix with the singular values
$\gamma_1, \gamma_2, \ldots$
as diagonal entries.
The $p \times p$ matrix $$S = \frac{1}{n} X^t X =
\frac{1}{n} V \Gamma^2 V^t $$ is the sample covariance
matrix of markers. The eigenvectors $\mbf{v}_1, \mbf{v}_2 \ldots$
are called {\em principal components}.
(If the columns of $X$ are furthermore
normalized to have standard deviation 1, then $S$ is the sample
correlation matrix of markers.)
In population genetics, 
Cavalli-Sforza and others
compute the dual $n \times n$ matrix 
 $$ H = X X^t =   U \Gamma^2 U^t
,$$ and use the rescaled eigenvectors of $H$ as coordinates of subject $i$,
\be (\lambda_1^{1/2} \mbf{u}_1(i),  \ldots, \lambda_d^{1/2} \mbf{u}_d(i)) 
\label{eq:eigen}), \ee
where $\lambda_j=\gamma_j^2$ and $\lambda_1 \geq \lambda_2 \geq \ldots$.
 Geometrically, this corresponds to projecting 
the data $\mbf{x}_i$ onto the affine hyperplane spanned by the first $d$ principal components,
i.e. computing the projection indices or principal component scores 
$(\mbf{x}_i \cdot \mbf{v}_1,  \ldots,   \mbf{x}_i \cdot \mbf{v}_d)$. 
 Typically, eigenvectors that correspond to large eigenvalues reveal the 
most important dimensions of ancestry.

The matrix $H$ 
is often referred
to as the ``covariance matrix of individuals'' but this is a bit of a
misnomer.
 In fact, some of the intuition behind the eigenmap method
  comes from thinking of
$H$ as an {\em inner product matrix} or Gram matrix. 
In multivariate statistics, the method of mapping data with principal
component scores is known as classical multidimensional
scaling. \cite{gower} made explicit the connection between classical
MDS and PCA, and demonstrated that the principal components can be
extracted from the inner product matrix $H$. The approach is also
directly related to kernel PCA~\citep{Scholkopf98} where all
computations are expressed in terms of $H$.

One can show that principal component mapping
solves a particular optimization problem with an
associated distance metric~\citep{torgerson,mardia78}. Refer to the centered data matrix as a
feature matrix $X$ where 
the $i$th row $\mbf{x}_i=(z_{i1}-\bar{z}_1, \ldots, z_{ip}-\bar{z}_p)$
is the ``feature vector'' of the $i$:th individual. In the
normalized case, the corresponding vector is
$\mbf{x}_i=(\frac{z_{i1}-\bar{z}_1}{s_1}, \ldots,
\frac{x_{ip}-\bar{z}_p}{s_p})$, where $\bar{z}_j$ and $s_j$,
respectively, are the sample mean and sample standard deviation of
variable (marker) $j$. 
 The matrix $H$
is a positive semi-definite (PSD) 
matrix, where element 
$h_{ij}= \mbf{x}_i \cdot \mbf{x}_j$ 
reflects the similarity between individuals $i$ and $j$. We will refer 
to $XX^t$ as the {\em kernel} of the PC map.
The main point is that the matrix $H$ induces a natural 
 Euclidean distance between individuals. We denote this Euclidean distance between the 
 $i^{th}$ and $j^{th}$ individuals as $m(i,j)$, where:
\be m(i,j)^2 \equiv   h_{ii} + h_{jj} - 2h_{ij} =  \|\mbf{x}_i-\mbf{x}_j\|^2  . 
\label{eq:metric}\ee

Consider a
low-dimensional representation $\Phi_d(i)=(\phi_1(i), \ldots,
\phi_d(i))$ of individuals $i=1,\ldots,n$, where the
dimension $d < p$. Define squared distances
$\widehat{m}(i,j)^2=\|\Phi_d(i)-\Phi_d(j)\|^2$ for this configuration.
  To measure the discrepancy between the
full- and low-dimensional space, let
$\delta=\sum_{i,j}(m(i,j)^2-\widehat{m}(i,j)^2)$.  This quantity is
minimized over all $d$-dimensional configurations by the top $d$
eigenvectors of $H$, weighted by the square root of the eigenvalues (Eq.~\ref{eq:eigen}); see Theorem 14.4.1 in \citep{mardia}.
  Thus principal component mapping
is a form of metric multidimensional scaling. It provides the
optimal embedding if the goal is to preserve the squared (pairwise) Euclidean distances
$m(i,j)^2$ induced by $H=XX^t$.

MDS was originally developed by psychometricians to visualize dissimilarity data~\citep{torgerson}.
The downside of using PCA for a quantitative analysis is that the
associated metric is highly sensitive to outliers, which diminishes
its ability to capture the major dimensions of ancestry. Our goal in
this paper is to develop a spectral embedding scheme that is less
sensitive to outliers and that is better, in many settings, at
clustering observations similar in ancestry.  We note that the choice
of eigenmap is not unique: {\em Any} positive semi-definite matrix $H$
defines a low-dimensional embedding and associated distance metric
according to Equations~\ref{eq:eigen} and~\ref{eq:metric}.  Hence, we
will use the general framework of MDS and principal component maps but
introduce a different kernel for improved
performance. Below we give some motivation for the modified kernel and
describe its main properties from the point of view of spectral graph
theory and spectral clustering.

 \subsection*{Spectral clustering and Laplacian eigenmaps}
Spectral clustering techniques~\citep{luxburg} use the spectrum 
of  the similarity matrix of the data to perform dimensionality reduction for 
clustering in fewer dimensions.
These methods are more flexible than
clustering algorithms that group data directly in the given coordinate system.
Spectral
clustering has not been, heretofore, fully explored in the context of
a large number of independent genotypes, such as is typically obtained
in genome-wide association studies. 
In the framework of spectral clustering, the decomposition of $XX^t$ in PCA
corresponds to an un-normalized clustering scheme.  Such schemes tend
to 
return embeddings where the principle axes separate outliers from
the bulk of the data.  On the other hand, an embedding based on 
a normalized data similarity matrix 
identifies directions with more balanced clusters.

To introduce the topic, we require the language of graph theory.
For a group of $n$ subjects, define a graph $G$ where $\{1, 2, \dots,n
\}$ is the vertex set (comprised of subjects in the study).  The
graph $G$ can be associated with a weight matrix $W$, that reflects
the strength of the connections between pairs of similar subjects: the higher
the value of the entry $w_{ij}$, the stronger the connection between
the pair $(i,j)$. Edges that are not connected have weight 0. There is flexibility in the choice of weights and 
there are many ways one can 
incorporate application- or data-specific information. The only condition on the matrix W is that it is symmetric with non-negative entries.

Laplacian eigenmaps \citep{belkin} find a new representation of the data 
by decomposing the so-called graph Laplacian --- a discrete version of the Laplace operator 
on a graph.
 Motivated by MDS, we consider a rescaled parameter-free variation of Laplacian eigenmaps. A similar approach 
 is used in diffusion maps~\citep{Coifman_PNAS1} and Euclidean commute time (ECT) maps~\citep{Fouss:EtAl:2007}; both of these methods are MDS-based and lead to Laplacian eigenmaps with rescaled eigenvectors\footnote{We have here chosen a spectral transform that is close to the original PC map but it is straight-forward to associate the kernel with a diffusion or ECT metric.}.
  
The Laplacian matrix
$L$ of a weighted graph $G$ is defined by
\[L(i,j)=
\left\{ \begin{array}{l}
-w_{ij}, \mbox{ if } \> i\neq j,\\
d_i-w_{ii}, \mbox{ if } i=j,
 \end{array} \right.\]
where $d_i= \sum_j w_{ij}$ is the so-called 
degree of vertex $i$. 
In matrix form,
\[L=D-W,\] where $D=diag(d_1, \dots, d_n)$ is a diagonal 
matrix.  The normalized graph Laplacian is 
a matrix defined as
\[ \mathcal{L}=D^{-1/2}LD^{-1/2}. \]  

A popular choice for weights is $w_{ij} = \exp(-\| \mbf{x}_i -  \mbf{x}_j\|^2/2 \sigma^2)$,
where the parameter $\sigma$ controls the size of local neighborhoods in the graph. 
Here we instead use a simple transformation of the (global) PCA kernel with no tuning 
parameters; in Discussion we later suggest a local kernel based on identity-by-state (IBS) sharing
for biallelic data. The main point is that one can choose a weight matrix suited for the particular application.
 Entries in the matrix $XX^t$ 
measure the similarity between subjects, making it a good candidate for
a weight matrix on a fully connected graph: the larger the entry for a pair $(i,j)$,
the stronger the connection between the subjects within the pair.
We define the weights as
\[ w_{ij}= \left\{ \begin{array}{l}
  \sqrt{\mbf{x}_i \cdot \mbf{x}_j}, \mbox{ if } \mbf{x}_i \cdot \mbf{x}_j \geq 0,\\
    0, \mbox{ otherwise.} \end{array} \right. \] 
Directly thresholding $XX^t$ guarantees non-negative weights but creates a skewed 
distribution of weights. To address this problem, we have added 
a square-root transformation for more symmetric weight distributions. This 
transformation also adds to the robustness to outliers.

Let $\nu_i$ and $\mbf{u}_i$ be the eigenvalues and eigenvectors of
 $\mathcal{L}$.  Let $\lambda_i =\max\{0,1-\nu_i\}$. We replace the PCA kernel $XX^t$ with $(I-\mathcal{L})_+$, where $I$ is the
identity matrix and $(I-\mathcal{L})_+ \equiv \sum_i \lambda_i \mbf{u}_i \mbf{u}_i^t$ 
is a positive semi-definite approximation of $I-\mathcal{L}$.
We then map the the $i$'th subject into a
lower-dimensional space according to Eq.~\ref{eq:eigen}.
 In embeddings, we often do not display the the first eigenvector 
$\mbf{u}_1$ associated with the eigenvalue $\lambda_1=1$, as this vector 
only reflects the square root of the degrees of the nodes.

In Results, we show that
estimating the ancestry from the eigenvectors of $\mathcal L$ (which
are the same as the eigenvectors of $I-\mathcal L$) leads to more
meaningful clusters than ancestry estimated directly from $XX^t$. Some
intuition as to why this is the case can be gained by relating
eigenmaps to spectral clustering and ``graph cuts''.
In graph-theoretic language, the goal of clustering is to find a
partition of the graph so that the connections between different
groups have low weight and the connections within a group have high
weight. For two disjoint sets $A$ and $B$ of a graph, the cut across
the groups is defined as $cut(A,B)=\sum_{i \in A, j \in B}
w_{ij}$. Finding the partition with the minimum cut is a well-studied
problem; however, as noted for example by Shi and Malik [1997] the
minimum cut criterion favors separating individual vertices or
``outliers'' from the rest of the graph.  The normalized cut approach
by Shi and Malik circumvents this problem by incorporating the volume
or weight of the edges of a set into a normalized cost function
$Ncut(A,B)=\frac{cut(A,B)}{vol(A)} + \frac{cut(A,B)}{vol(B)},$
where
$vol(A)=\sum_{i\in A}d_i$ and $vol(B)=\sum_{i\in B}d_i$.  This cost
function is large when the set $A$ or $B$ is small.  Our SpectralGEM
algorithm (below) exploits the fact that the top eigenvectors of the graph
Laplacian provide an approximate solution to the Ncut minimization
problem; see Shi and Malik for details. 
Smartpca \citep{patterson} and standard GEM \citep{t1d}, on
the other hand, are biased towards embeddings that favor small and 
tight clusters in the data.

\subsection*{Number of dimensions via eigengap heuristic}
For principal component maps, one can base a formal test for the number
of significant dimensions on theoretical results
concerning the Tracy-Widom distribution of eigenvalues of a covariance matrix in
the null case
\citep{patterson,johnstone}. 
Tracy-Widom theory does not extend to the eigenvalues of the graph
Laplacian where matrix elements are correlated. Instead we introduce a different approach,
known as the eigengap heuristic, based on the difference in magnitude
between successive eigenvalues.

The graph Laplacian has several properties that make it useful 
for cluster analysis. Both its eigenvalues and eigenvectors 
reflect the connectivity of the data. Consider, for example, the 
normalized graph Laplacian where the sample consists of $d$ distinct clusters. 
Sort the eigenvalues 
$0=\nu_1 \leq \nu_2 \leq \ldots \leq \nu_n$ of $\mathcal{L}$ 
in ascending order.
The matrix $\mathcal{L}$ has several key properties 
\citep{chung}: 
(i) The number $d$ of eigenvalues equal to $0$ is the number 
of connected components $S_1, \ldots, S_d$ of the graph.
(ii) The first positive eigenvalue $\nu_{d+1}$
reflects the cohesiveness 
of the individual components;
 the larger the eigenvalue $\nu_{d+1}$ the more cohesive the clusters.
 (iii) The eigenspace of $0$ (i.e., the vectors corresponding to eigenvalues equal to 0) is spanned
by the rescaled indicator vectors $D^{1/2} \mathbf{1}_{S_k}$, 
where $ \mathbf{1}_{S_k}=1$ if $i\in S_k$, and $\mathbf{1}_{S_k}=0$ otherwise.
It follows from (iii) that for the ideal case where we have $d$ 
completely separate populations (and the node degrees are similar), 
individuals from the same population map 
into the same point in an embedding defined by the $d$ first eigenvectors 
of $\mathcal{L}$. For example, for $d=3$ populations and $n=6$ individuals, 
the $n \times d$ embedding matrix could have the form
$$
    U = [D^{1/2} \mathbf{1}_{S_1},\ D^{1/2} \mathbf{1}_{S_2},\ D^{1/2} \mathbf{1}_{S_3} ]
   =  \left( \begin{array}{ccc}
    \sqrt{d_1} & 0 & 0\\
     \sqrt{d_2} & 0 & 0\\
     \sqrt{d_3} & 0 & 0\\
    0 &  \sqrt{d_4} & 0 \\
    0 &  \sqrt{d_5} & 0\\
    0 & 0 &  \sqrt{d_6}
    
     \end{array} \right)
    \approx \left( \begin{array}{ccc}
    1 & 0 & 0\\
    1 & 0 & 0\\
    1 & 0 & 0\\
    0 & 1 & 0 \\
    0 & 1 & 0\\
    0 & 0 & 1
     \end{array} \right).
$$
The rows of $U$ define the new representation of the $n$ individuals. 
Applying $k$-means to the rows finds the clusters trivially 
without the additional assumption on the node degrees, if one as 
in the clustering algorithm by~\cite{njw} first re-normalizes the rows of $U$ to norm $1$, 
or if one according to~\cite{shi}  computes eigenvectors 
of the graph Laplacian $I-D^{-1}W$ instead of the symmetric Laplacian 
$I-D^{-1/2}WD^{-1/2}$.
 
In a more realistic situation the between-cluster similarity will 
rarely be exactly $0$ and all components of the graph will be
connected. Nevertheless, if the clusters are distinct, we may still use 
the eigenvalues of the graph Laplacian to determine the number of 
significant dimensions. Heuristically, choose the number $d$ of significant 
eigenvectors such that the 
eigengaps $\delta_i = | \nu_{i+1}-\nu_{i}|$ 
are small for $i<d$ but the eigengap $\delta_d$ is large. 
One can justify such an approach
with an argument from perturbation analysis \citep{stewart}. The idea is that the matrix 
$\mathcal{L}$ for the genetic data is a perturbed version of the 
ideal matrix for
$d$ disconnected clusters. If the perturbation is not too large 
and the ``non-null'' 
eigengap $\delta_d$ 
is large, 
the subspace spanned by the first $d$ eigenvectors will be close to the 
subspace defined by the ideal indicator vectors and a 
spectral clustering algorithm will separate the individual clusters well. 
The question then becomes: How do we decide 
whether an eigengap is significant (non-null)?

In this work, we propose
 a practical scheme for estimating the number of significant 
 eigenvectors for genetic ancestry
that is based on the eigengap heuristic and hypothesis testing. 
By simulation, we generate homogeneous data without population structure
and study the distribution of eigengaps for the normalized graph Laplacian. 
Because there is only one population, the first eigengap $\delta_1$ is large. 
We are interested in the first null eigengap, specifically 
the difference $\delta_2=|\nu_3-\nu_2|$ between the 2nd 
and 3rd eigenvalues (note that $\nu_1$ is always $0$). 
If the data are homogeneous, this difference is relatively small. 
 Based on our simulation 
results, we approximate the upper bound for the null eigengap with 
the 99th quantile of the sampling distribution as a function of the number of subjects $n$ 
and the number of SNPs $p$. 
 In the eigenvector representation, we choose
the dimension $d$ according to
\[ d = \max \{i; \delta_i> f(n,p) \},\]
where $f(n,p) = -0.00016 +2.7/n + 2.3/p$ is the empirical expression for the 99th quantile. 
For most applications, we have that $p \gg n$ and $f(n,p) \approx 2.7/n$.

\subsection*{Controlling for ancestry in association studies}
\begin{figure}
\includegraphics[angle=0,width=7in]{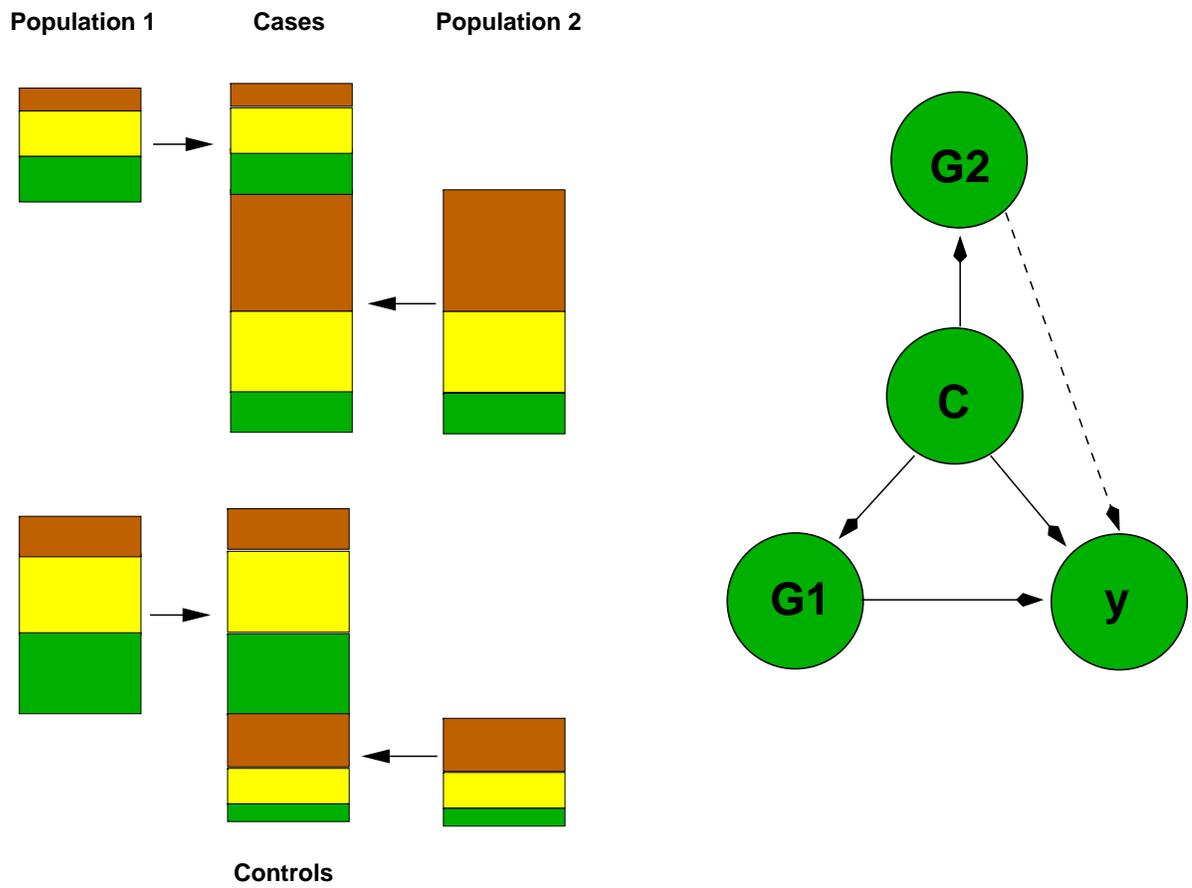}
\caption{Example of population stratification
due to both disease prevalence and allele frequencies 
varying by ancestry. See text for details.} 
\label{fig:pop}
\end{figure}

Due to demographic, biological and random forces, genetic variants 
differ in allele frequency in populations around the world.  
A case-control study could be susceptible to {\em population stratification},
a form of confounding by ancestry, when such variation is correlated with 
other unknown risk factors. \myfigref{fig:pop}
shows an example of population stratification. We wish to test 
the association between candidate SNPs and the outcome (Y) of a disease. 
In the example, the genotype distributions for Population 1 and 2 are different, 
as illustrated by the different proportions of red, yellow and green.
In addition, there are more cases (Y=1) from Population 2 than 1, and more controls (Y=0) from Population 1 than 2.
 Let G1 and G2, respectively, be the genotypes of a causal versus a non-causal SNP.
The arrow from G1 to Y in the graph to the right 
indicates a causal association. There is no causal association 
between G2 and Y but the two variables are indirectly associated, 
as indicated by the dotted line, through ancestry (C). Ancestry 
is here a ``confounder'' as it is {\em both} associated with allele frequency 
{\em and} disease prevalence conditional on genotype; it 
distorts the assessment of the direct relationship between G2 and Y and 
decreases the power of the study.

Statistical techniques to control spurious findings 
include stratification 
by the Cochran-Mantel-Haenszel method, 
regression and matching~\cite{rosenbaum}.
These approaches assume that 
the key confounding factors have been identified, 
and that at each distinct level of the confounders, 
the observed genotype is 
independent of the case and control status. 
In this work, we estimate confounding ancestry by an eigenanalysis 
(PCA or spectral graph) of a panel of reference SNPs. Under the additional assumption 
that the interaction between ancestry and the genotype of the candidate SNPs
is neglible, we  
compare different techniques of controlling for ancestry.

The most straight-forward strategy to correct for stratification is to embed the data
using the inferred axes of variation and divide the population into 
$K$ groups or strata that are homogeneous with respect to ancestry.
The {\em Cochran-Mantel-Haenszel (CMH) method} represents the data as a series 
of $K$ contingency tables. 
One then performs a chi-squared test of the null hypothesis that the 
disease status is conditionally independent of the genotype in any given stratum.
The precision in sample estimates of the CMH test statistic is sensitive to 
 the sample size as well as the 
 balance of the marginals in the contingency table. This can be a problem if 
we have insufficient data or if 
cases and controls are sampled from different populations. 

An alternative approach is to use a {\em regression model} for the  
disease risk as a function of allele frequency. Effectively, regression models link
information from different strata by smoothness assumptions. Suppose 
that $x$ is the observed allele count (0, 1 or 2) of the candidate SNP, and that 
the eigenmap coordinates of an individual is given by  $\phi_1, \ldots, \phi_d$. 
Assign $Y=1$ to cases and $Y=0$ to controls and let
$q=P(Y=1|x,\phi_1,\ldots,\phi_d)$.  For a logistic regression model
$$\log\left(\frac{q}{1-q}\right) = \beta x + b_1 \phi_1 + \ldots + b_d \phi_d \ ,$$
the regression parameter $\beta$ can be interpreted as the increase in the log odds
of disease risk per unit increase in $x$, holding all other risk variables 
in the model constant. Thus, the null hypothesis  $H_0: \beta=0$ is equivalent to 
independence of disease and SNP genotype after adjusting 
for ancestry.

A third common strategy to control for confounding is to produce 
a fine-scale stratification of ancestry by {\em matching}.  
Here we use a matching scheme introduced in an earlier paper~\citep{t1d}.
The starting point is to estimate ancestry using an eigenanalysis (PCA for ``GEM'' and 
the spectral graph approach for ``SpectralGEM''). 
Cases and controls are matched with respect to the Euclidean metric 
in this coordinate system; hence, the relevance of an MDS interpretation with an explicitly defined metric. Finally, we perform 
conditional logistic regression for the matched data.

\subsection*{Algorithm for SpectralR and SpectralGEM}
Algorithm~\ref{alg:spectral_GEM} summarizes the two related avenues
that use the spectral graph approach to control for genetic ancestry:
SpectralR (for Regression) and 
SpectralGEM (for GEnetic Matching). 

There are many possible variations 
of the algorithm. In particular, the normalization and rescaling in 
Steps 3 and 7 can be adapted to 
the clustering algorithms by Shi-Malik and Ng-Jordan-Weiss. One can also redefine 
 the weight matrix in Step 2 to model different structure in the genetic data.
\begin{algorithm}
\caption{SpectralR and SpectralGEM}
\label{alg:spectral_GEM}
\begin{algorithmic}[1]
\STATE Center and scale the allele counts. Let $\mbf{x}_i$ be the genetic information for subject $i$.
\STATE Compute weight matrix $W$ where $w_{ij}=(\max \{{\mbf{x}_i \cdot \mbf{x}_j},0\})^{1/2}$.
\STATE Compute the normalized Laplacian matrix $\mathcal{L}=I - D^{-1/2}WD^{-1/2}$.
\STATE Find the eigenvalues  $\nu_i$ and eigenvectors $\mbf{u}_i$ of $\mathcal{L}$.
\STATE Define the PSD matrix $H = (I - \mathcal{L})_{+}$ with eigenvalues  
$\lambda_i = \max\{0, 1-\nu_i\}$ and eigenvectors  $\mbf{u}_i$. This is the {\em kernel} of our map.
\STATE Determine the number of significant dimensions $d$ in the eigenvector representation
\[ d = \max \{i; \delta_i> -0.00016 +2.7/n + 2.3/p \},\]
\STATE Let $\Phi_d(i) = (\lambda_1^{1/2} \mbf{u}_1(i), \dots, \lambda_d^{1/2} \mbf{u}_d(i))$ 
be the new representation of subject $i$.

\STATE {\bf For regression (SpectralR):}
\STATE  \hspace{.2in}  Perform logistic regression with the the $d$ eigenmap coordinates 
and the allele count of the candidate SNP as covariates.
\STATE  \hspace{.2in}  Compute p-values for the Wald test of no association between disease and SNP genotype.

\STATE {\bf For genetic matching (SpectralGEM):}
\STATE \hspace{.2in} Compute the distance between subjects $i$ and $j$ using $\| \Phi_d(i) - \Phi_d(j) \|$.
\STATE \hspace{.2in} Find homogeneous clusters of individuals via Ward's $k$-means algorithm \citep{t1d}.
\STATE \hspace{.2in} Rescale the data as described in the GEM algorithm~\citep{t1d}.
 \STATE \hspace{.2in} Remove unmatchable subjects prior to
  analysis.  
\STATE \hspace{.2in} Recompute the eigenmap. Match cases and controls in $d$ dimensions.
\STATE \hspace{.2in} {Perform conditional logistic regression and compute p-values for the 
Wald test.}
\end{algorithmic}
\end{algorithm}

 \section*{Analysis of Data}

 A large number of subjects participating in multiple studies
 throughout the world have been assimilated into a freely available
 database known as POPRES \citep{nelson}.  Data consists of genotypes
 from a genome-wide 500,000 single-nucleotide polymorphism panel. This
 project includes subjects of African American, E. Asian,
 Asian-Indian, Mexican, and European origin.  We use these data to
 assess performance of spectral embeddings.  For more detailed
 analyses of these data see \citep{lee}.

These data are challenging because of the 
disproportionate representation of individuals of European ancestry
combined with individuals from  multiple continents.  To obtain
results more in keeping with knowledge about population demographics, 
\cite{nelson} supplement POPRES with 207 unrelated subjects
from the four core HapMap samples. In addition, to overcome problems
due to the dominant number of samples of European ancestry, they remove
889 and 175 individuals from the Swiss and U.K. samples, respectively.
Because PCA is sensitive to outliers, they perform a careful search
for outliers, exploring various subsets of the data iteratively.
After making these adjustments they obtain an excellent description of
the ancestry of those individuals in the remaining sample, detecting
seven informative axes of variation that highlight important features of the genetic
structure of diverse populations. 
When analysis is restricted to individuals of
European ancestry, PCA works very well \citep{novembre}.  Direct
application of the approach to the full POPRES data leads to much less useful insights
as we show below.

    \subsection*{Data Analysis of POPRES}

   Demographic records in POPRES include the individual's
 country of origin and that of his/her parents and grandparents.
 After quality control the data included 2955 individuals of European
 ancestry and 346 African Americans, 49 E. Asians, 329 Asian-Indians,
 and 82 Mexicans.  From a sample of nearly 500,000 SNPs we focus on
 21,743 SNPs for in depth analysis.  These SNPs were chosen because
 they are not rare (minor allele frequency $\ge .05$), and have a low
 missingness rate ($\le .01$).  Each pair is separated by at least 10
 KB with squared correlation of 0.04 or less.  

{\it Outlier Dataset.} It is well known that outliers can interfere with
discovery of the key eigenvectors and increase the number of significant
dimensions discovered with PCA.  To illustrate the effect of outliers we
created a subsample from POPRES including 580 Europeans (all
self-identified Italian and British subjects), 1 African American, 1
E. Asian, 1 Indian and 1 Mexican.  Smartpca removes the 4 outliers
prior to analysis and discovers 2 significant dimensions of ancestry.
If the outliers are retained, 5 dimensions are significant. The first
two eigenvectors separate the Italian and British samples and
highlight normal variability within these samples. Ancestry vectors
3-5 isolate the outliers from the majority of the data, but otherwise
convey little information concerning ancestry.

With SpectralGEM, leaving the outliers in the data has no impact.
The method identified 2 significant dimensions that are nearly
identical to those discovered by PCA.  In our cluster
analysis we identified 4 homogeneous clusters: 1 British cluster, 2
Italian clusters, and 1 small cluster that includes the outliers and
6 unusual subjects from the remaining sample.

{\it Cluster Dataset.} The ancestral composition of samples for genome-wide
association studies can be highly variable. To mimic a typical
situation we created a subsample from POPRES including 832 Europeans
(all self-identified British, Italian, Spanish and Portuguese
subjects), 100 African Americans and 100 Asian-Indians.

\begin{figure}
\includegraphics[angle=0,width=7in]{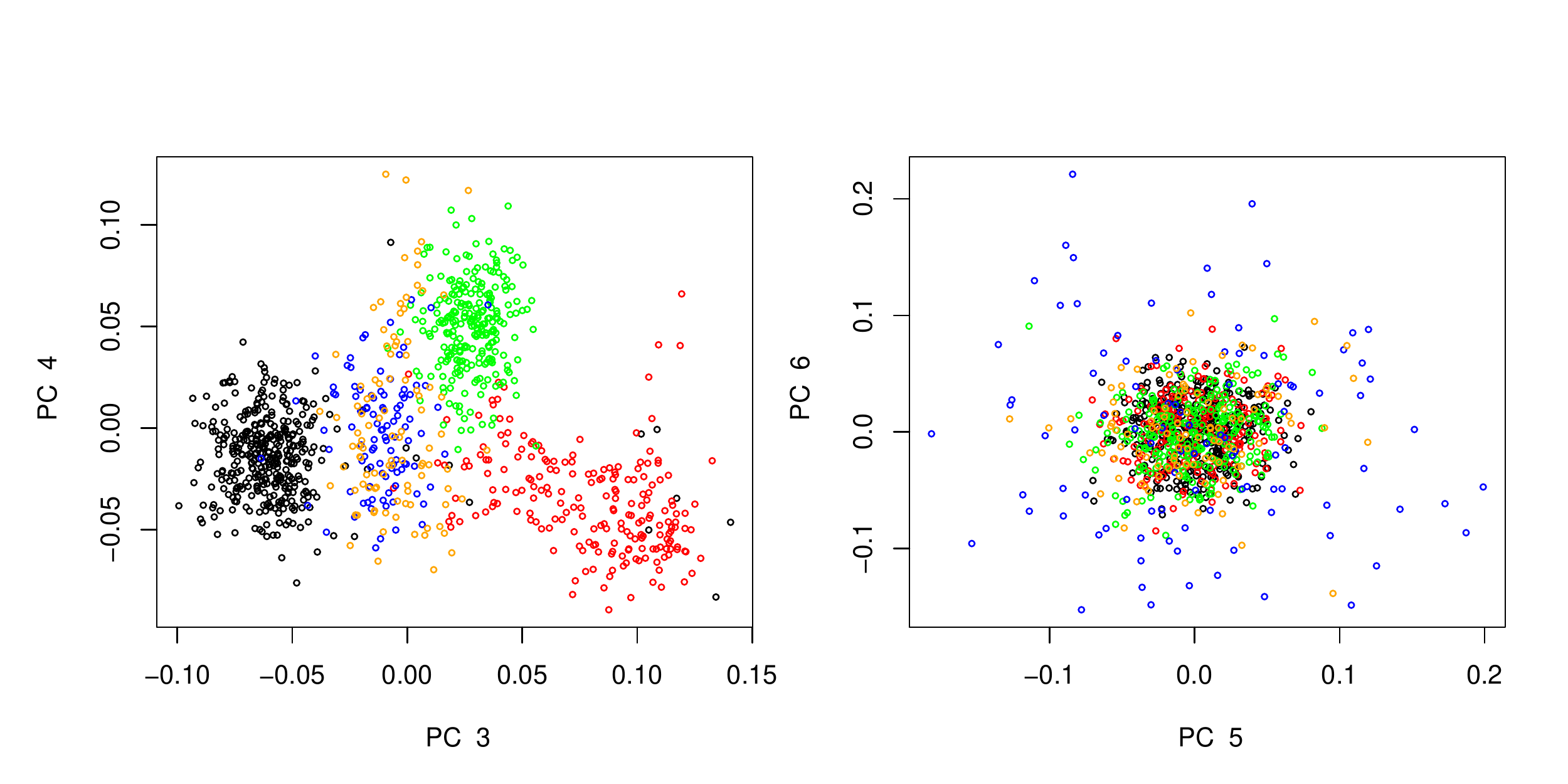}
\caption{Principal components 3-6 for data from the Cluster
Dataset.  PC 1 and PC 2 are quite similar to the eigenvectors shown in
Fig. 4.  Subjects are self-identified as U.K. (black), Italian (red),
Iberian Peninsula (green), African American (blue), Indian (orange)}
\label{fig:PCA_cluster}
\end{figure}

Using smartpca, 7 dimensions of ancestry are significant.  The first 2
eigenvectors separate the continental samples.  The third and fourth
eigenvectors separate the Europeans roughly into three domains
( \myfigref{fig:PCA_cluster}).  The three European populations form three clusters, but 
they are not completely delineated.  The other continental
groups generate considerable noise near the center of the plot. The
remaining 3 significant dimensions reveal little structure of
interest.

\begin{figure}
\includegraphics[angle=0,width=7in]{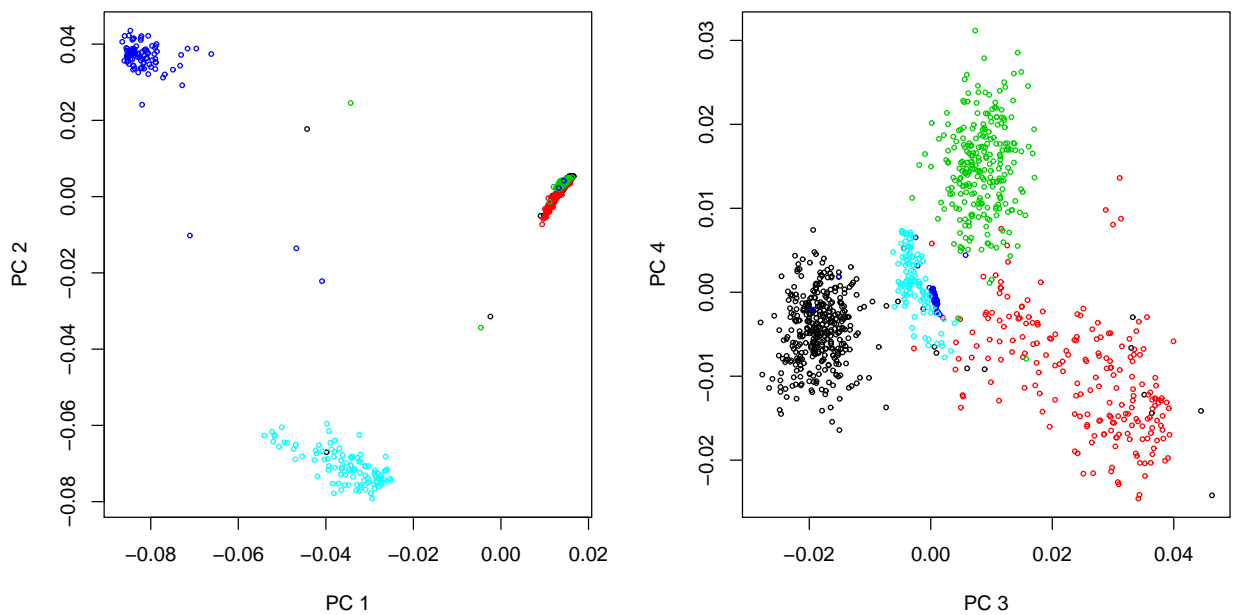}
 \caption{Non-trivial eigenvectors (EV) from the spectral graph approach
  for the Cluster Dataset.  Subjects are self-identified as U.K. (black), Italian
  (red), Iberian Peninsula (green), African American (blue), Indian
  (orange).}
\label{fig:GEM_cluster}
\end{figure}

Using SpectralGEM 4 dimensions are significant (\myfigref{fig:GEM_cluster}). The first
two dimensions separate the continental clusters. In
the third and fourth dimensions, the European clusters separate more
distinctly than they did for PCA.  For these higher dimensions, the
samples from other continents plot near to the origin, creating a
cleaner picture of ancestry.  Six homogeneous clusters are discovered,
3 European clusters, an African American cluster and 2 Indian
clusters.

{\it Full Dataset.} For the greatest challenge we analyze the full
POPRES sample.  Smartpca's 6 standard deviation outlier rule removes
141 outliers, including all of the E. Asian and Mexican samples. If
these ``outliers'' were retained, PCA finds 12 significant dimensions:
the first 4 dimensions separate the 5 continental populations
(African, European, Latin American, E. Asian and S. Asian).  Other
eigenvectors are difficult to interpret.  Moreover, based on this
embedding, Ward's clustering algorithm failed to converge; thus no
sensible clustering by ancestry could be obtained.

With SpectralGEM no outliers are removed
prior to analysis.  The number of significant dimensions of ancestry
is 8.  The first 4 dimensions separate the major continental samples;
the remaining dimensions separate the European sample into smaller
homogeneous clusters.

Applying the clustering algorithm based on this eight dimensional
embedding we discover 16 clusters and 3 outliers.  Four of these
clusters group the African American, E. Asian, Indian and Mexican
samples, so that greater than 99\% of the
subjects in a cluster self-identified as that ancestry, and only a
handful of subjects who self-identified as one of those four
ancestries fall outside of the appropriate cluster.  

The remaining 12 clusters separate the individuals of European
ancestry.  For ease of interpretation, we removed the samples obtained
from Australia, Canada, and the U.S., and focus our validation on 2302
European samples, which can be more successfully categorized by
ancestry based on geographic origin.  These individuals were
classified to one of the 34 European countries represented in the
database (Table 1). Sample sizes varied greatly across countries.
Seven countries had samples of size 60 or more.  Countries with
smaller samples were combined to create composite country groupings
based on region; see Table 1 for definition of country groupings.

\begin{figure}
\includegraphics[angle=0,width=7in]{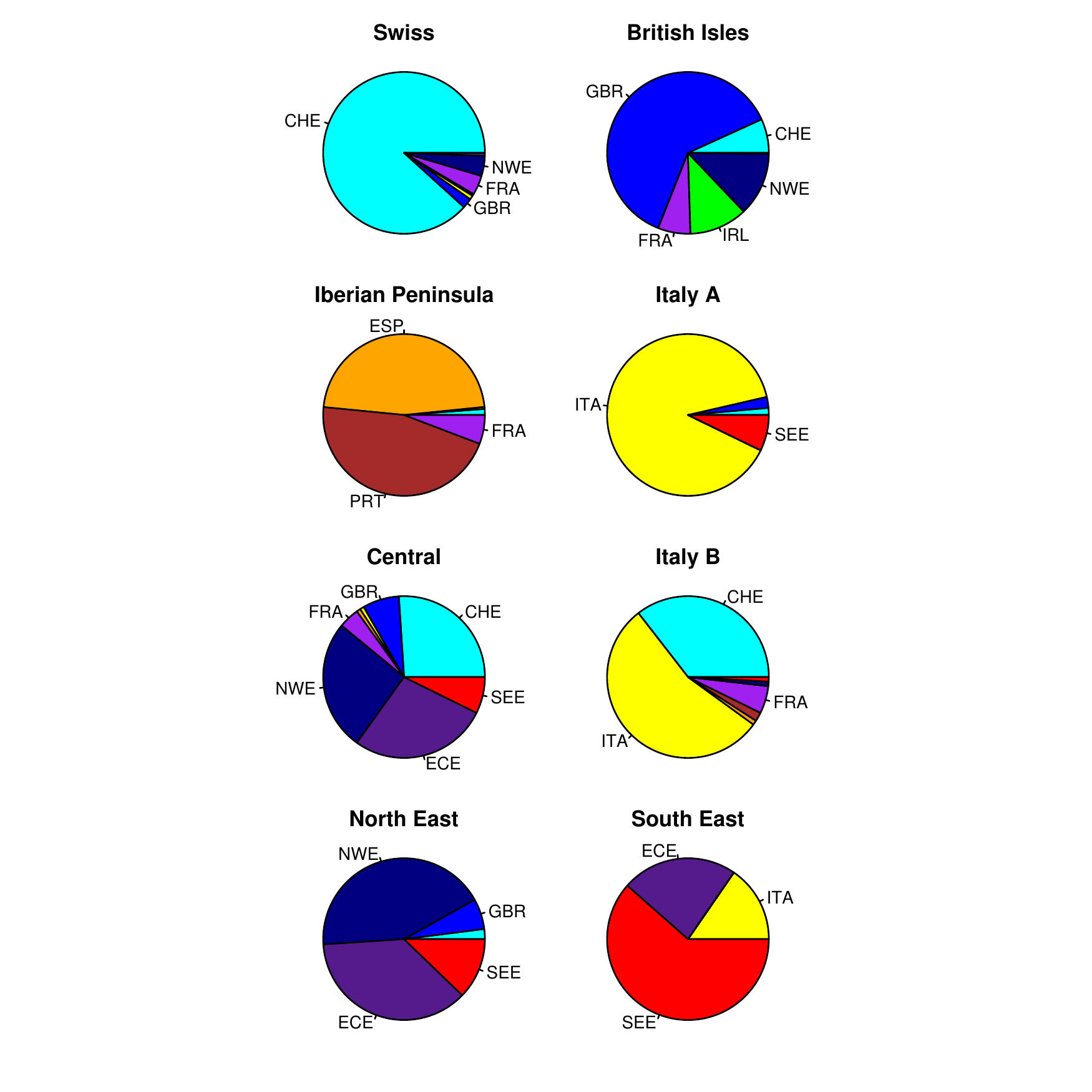}
\caption{Country membership by cluster for the Full
Dataset. Cluster labels and country groupings are defined in Table
I. Cluster labels were derived from the majority country or country
grouping membership.}
\label{fig:country_membership}
\end{figure}

By using Ward's clustering algorithm based on the spectral embedding,
all but 81 of the European sample were clustered into one of 8
relatively large European clusters (labeled A-H, Table 1).  \myfigref{fig:country_membership} 
illustrates the conditional probability of country grouping given
cluster. Clusters tend to consist of individuals sampled from a
common ancestry.  Labeling the resulting clusters in  \myfigref{fig:country_membership}  by
the primary source of their membership highlights the results: (A)
Swiss, (B) British Isles, (C) Iberian Peninsula, (D) Italian A, (E)
Central, (F) Italian B, (G) North East, and (H) South East. The
remaining four small clusters show a diversity of membership and are
simply labeled I, J, K, and L.  Cluster L has only 7 members who could
be classified by European country of origin.

\begin{figure}
\includegraphics[angle=0,width=7in]{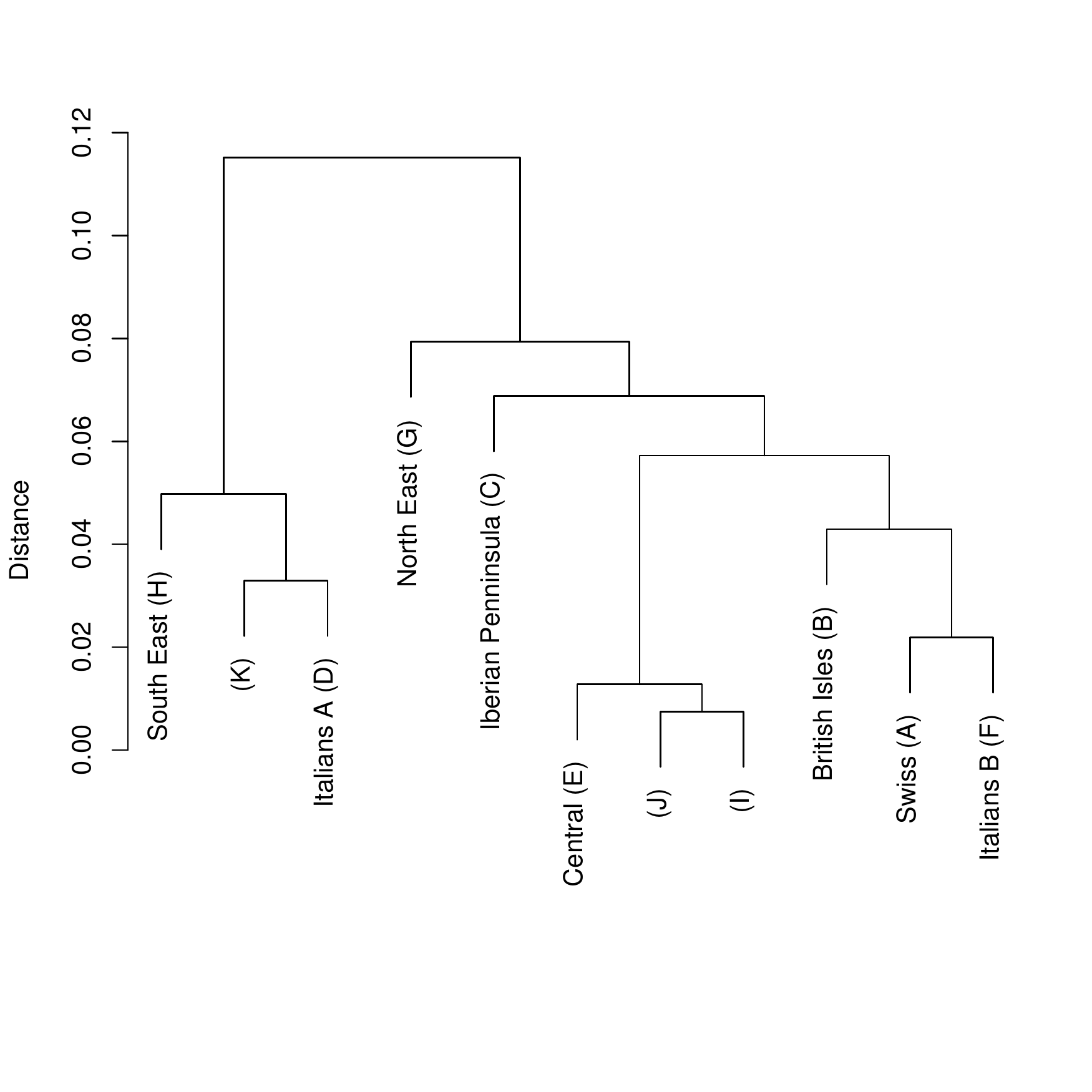}
\caption{Dendrogram for European clusters from the Full Dataset.}
\label{fig:dendrogram}
\end{figure}

A dendrogram displays the relationships between clusters (\myfigref{fig:dendrogram}).
For instance, it appears that the Italian A and B clusters represent
Southern and Northern Italy, respectively.  Clusters I and J are
similar to the Central cluster, while Cluster K represents a more
Southern ancestry.

\subsection*{Simulations for association}
To compare smartpca with SpectralGEM and SpectralR using the POPRES
data it is necessary to create cases ($Y=1$) and controls ($Y=0$) from
this undifferentiated sample.  Disease prevalence often varies by
ancestry due to genetic, environmental and cultural differences. To
simulate a realistic case-control sample we wish to mimic this
feature.  We use cluster membership, $C=k$, $k=1,\ldots,K$, as a proxy
for ancestry and assign cases differentially to clusters.  In our
previous analysis we identified 16 clusters, 12 of European ancestry
and 4 of non-European ancestry.  For simplicity we reduce the number
of European clusters to 8 using the dendrogram and Table 1 to help
group the small clusters: K with D, and I, J and L with E.

To generate an association between $Y$ and $C$ we vary $P(Y=1|C=k)$ by
cluster. Within each cluster, case and control status is
assigned at random.  This creates a relationship between $Y$ and the
observed SNPs that is purely spurious.  Thus we can assess the Type I
error rate of smartpca and SpectralGEM to evaluate the efficacy of
the two approaches in removing confounding effects induced by
ancestry. 

To assess power we must generate SNPs associated with $Y$ using a
probability model.  To maintain as close a correspondence with the
observed data as possible, we simulate each causal SNP using the
baseline allele frequencies, $p_k,\,k=1,\ldots,12$, obtained from a
randomly chosen SNP in the data base.  For cluster $k$, when the
individual is a control the simulated genotype is 0, 1 or 2 with
probabilities $(1-p_k)^2, 2p_k(1-p_k)$ or $p_k^2$, respectively.  The
association is induced by imposing relative risk $R>1$ which
corresponds with the minor allele at a simulated causal locus.  Case
individuals are assigned genotype 0, 1 and 2 with probabilities
proportional to $(1-p_k)^2, 2Rp_k(1-p_k)$ and $R^2p_k^2$,
respectively.  We repeat this process to generate $M=1000$ SNPs
associated with $Y$.

We wish to compare two approaches for estimating ancestry (PCA and
spectral graph) and two approaches for controlling ancestry
(regression and matching).  Luca et al. (2008) conducted a thorough
comparison between regression and matching using eigenvectors derived
from PCA.  Here we focus on two key comparisons: (i) we control
confounding using regression and compare the efficacy of eigenvectors
estimated using PCA versus the spectral graph approach (Smartpca
versus SpectralR); and (ii) we estimate eigenvectors using the
spectral graph approach and compare efficacy of matching versus the
regression approach (SpectralGEM versus SpectralR).  Finally, we
compare all of these methods to the CMH approach which uses the
clusters as strata.

We perform the following experiment: randomly
sample half of the POPRES data; assign case and control status
differentially in clusters according the the model $P(Y|C)$; estimate
the eigenvectors using the two approaches based on the $p$ observed
SNPs; assess Type I error using the observed SNPs; generate $M$
causal SNPs; assess power using the simulated SNPs.  
From our previous analysis we know that all of the samples of 
Indian and Mexican ancestry are declared outliers using the 6 sd rule for outliers.
Most practitioners, however, would not discard entire
clusters of data.  Thus we do not remove
outliers in the simulation experiment.

We simulate a disease with differential sampling of cases from each
cluster to induce spurious association between $Y$ and the observed
genotypes.  This experiment is repeated for 5 scenarios (Table 2). In
Scenario 1, $P(Y=1|C=k)$ varies strongly by continent, but is
approximately constant within Europe.  In Scenarios 2 and 3 the tables
are reversed, with the variability most exaggerated within Europe.
This could occur in practice due to differential efforts to recruit
cases in different regions.  In Scenarios 3 and 4, $P(Y=1|C=k)$
approximately follows a gradient across Europe with high prevalence in northern
Europe and low prevalence in southern Europe. In Scenario 5,
$P(Y=1|C)=0$ for three of the small clusters to simulate a situation
where some controls were included for convenience, but no cases of
corresponding ancestry were included in the study.

All four approaches controlled rates of spurious association fairly
well compared to a standard test of association (Table 3).  
Overall, matching is slightly more conservative than regression (Table
3).  For Scenarios 1-4 this leads to a slight excess of control of
Type I errors. For Scenario 5, the advantages of matching come to the
fore.  When regions of the space have either no cases or no controls,
the regression approach is essentially extrapolating beyond the range
of the data.  This leads to an excess of false positives that can be much
more dramatic than shown in this simulation in practice 
\citep{t1d}. The matching approach has to have a minimum of one case
and one control per strata, hence it downweights samples that are
isolated by pulling them into the closest available strata.

For each scenario the number of significant eigenvectors was 6 or 7
using the spectral graph approach.  With PCA the number of dimensions 
was 16 or 17, i.e the method overestimates the number of 
important axes of variation. With respect to power, however, there 
is no penalty for using too many dimensions since the axes are orthogonal. 
This may explain why the power of smartpca was either equivalent or slightly higher than 
the power of SpectralR in our simulations.
Because
matching tends to be conservative, 
it was also not surprising to find that
the power of SpectralR was greater than SpectralGEM.  Finally, all of
these approaches exhibited greater power than the CMH test, suggesting
that control of ancestry is best done at the fine scale level of
strata formed by matching cases and controls than by conditioning on
the largest homogeneous strata as is done in the CMH test.


\section*{Discussion}

Mapping human genetic variation has long been a topic of keen
interest.  
\cite{cavalli} assimilated data from
populations sampled worldwide.  From this they created
PC maps displaying variation in allele frequencies that dovetailed
with existing theories about migration patterns, spread of innovations
such as agriculture, and selective sweeps of beneficial mutations.
Human genetic diversity is also crucial in determining the genetic
basis of complex disease; individuals are measured at large numbers of
genetic variants across the genome as part of the effort to discover
the variants that increase liability to complex diseases. Large,
genetically heterogeneous data sets are routinely analyzed for
genome-wide association studies. These samples exhibit complex
structure that can lead to spurious associations if differential
ancestry is not modeled~\citep{lander,pritchard2,devlin}.

While often successful in modelling the structure in data, PCA has
some notable weaknesses, as illustrated in our exploration of POPRES
\citep{nelson}.  In many settings the proposed spectral graph
approach, is more robust and flexible than PCA.  Moreover,
finding the hidden structure in human populations using a small number of
eigenvectors is inherently appealing.

A theory for the eigenvalue distribution of Laplacian matrices, analogous 
to the Tracy-Widom distribution for covariance matrices, is however not yet
available in the literature.  
Most of current results concern upper and lower bounds for the eigenvalues 
of the Laplacian~\citep{chung},  the distribution of all eigenvalues of the matrix as a
whole for random graphs with given expected degrees~\citep{Chung:EtAl:2003}, 
and rates of convergence and distributional limit theorems for the difference between 
the spectra of the random graph Laplacian $H_n$ and its limit $H$~\citep{koltchinskii,ShaweTaylor2002,
ShaweTaylor2005}.
At present, we rely on simulations of homogeneous populations in our work to
derive an approximation to the distribution of the key eigengap.

Furthermore, the weight matrix for the spectral graph implemented here was
motivated by two features: it is quite similar to the PCA kernel used
for ancestry analysis in genetics; and it does not require a tuning
parameter.  Nevertheless we expect that a local kernel with a tuning
parameter could work better. Because the features (SNPs) take on only
3 values, corresponding to three genotypes, the usual Gaussian kernel
is not immediately applicable. To circumvent this difficulty, a
natural choice that exploits the discrete nature of the data to
advantage is based on ``IBS sharing''.  For individuals $i$ and $j$, let
$s_{ij}$ be the fraction of alleles shared by the pair identical by
state across the panel of SNPs \citep{Weir}.  Define the
corresponding weight as $w_{ij}=\exp\{-(1-s_{ij})^2/\sigma^2\}$, with
tuning parameter $\sigma^2$.  Preliminary investigations suggest that this
kernel can discover the hierarchical clustering structure
often found in human populations, such as major continental clusters,
each made up of subclusters. Further study is required to develop a
data-dependent choice of the tuning parameter.

\vskip 1in
\noindent 
{\bf Acknowledgements.} This work was funded by NIH (grant MH057881) and 
ONR (grant N0014-08-1-0673).  

\vskip 1in
\noindent A public domain R language package SpectralGEM is available
from the CRAN website, as well as from
http://wpicr.wpic.pitt.edu/WPICCompGen/

\newpage

\bibliography{bibpaper}
\bibliographystyle{chicago}

\newpage
{\baselineskip 10pt
\scriptsize
\begin{center}
\begin{tabular}{lcrrrrrrrrrrrrr} \hline
Country & Subset & Count &\multicolumn {12}{c}{Cluster Label}\\
\hline
&&&A&B&C&D&E&F&G&H&I&J&K&L\\ \hline
Switzerland&CHE&1014&871&36&3&2&32&39&1&0&9&14&5&2\\
England&GBR&26&0&22&0&0&1&0&0&0&1&0&2&0\\
Scotland&GBR&5&0&5&0&0&0&0&0&0&0&0&0&0\\
United Kingdom&GBR&344&20&300&0&3&8&0&3&0&1&1&5&1\\
Italy&ITA&205&8&0&1&124&1&60&0&4&1&2&4&0\\
Spain&ESP&128&3&0&122&0&1&1&0&0&0&0&1&0\\
Portugal&PRT&124&1&0&119&0&0&2&0&0&0&0&0&0\\
France&FRA&108&39&34&15&0&5&6&0&0&3&2&3&1\\
Ireland&IRL&61&0&61&0&0&0&0&0&0&0&0&0&0\\
Belgium&NWE&45&21&19&0&0&3&0&0&0&1&1&0&0\\
Denmark&NWE&1&0&1&0&0&0&0&0&0&0&0&0&0\\
Finland&NWE&1&0&0&0&0&0&0&1&0&0&0&0&0\\
Germany&NWE&71&16&22&0&0&22&1&3&0&3&0&2&2\\
Latvia&NWE&1&0&0&0&0&0&0&1&0&0&0&0&0\\
Luxembourg&NWE&1&0&0&0&0&1&0&0&0&0&0&0&0\\
Netherlands&NWE&19&3&15&0&0&1&0&0&0&0&0&0&0\\
Norway&NWE&2&0&2&0&0&0&0&0&0&0&0&0&0\\
Poland&NWE&21&0&1&0&0&3&0&16&0&1&0&0&0\\
Sweden&NWE&10&0&7&0&0&2&0&0&0&0&1&0&0\\
Austria&ECE&13&3&1&0&0&6&0&0&0&2&0&0&1\\
Croatia&ECE&8&0&0&0&0&5&0&2&1&0&0&0&0\\
Czech&ECE&10&1&0&0&0&6&0&3&0&0&0&0&0\\
Hungary&ECE&18&0&0&0&0&10&0&4&1&2&1&0&0\\
Romania&ECE&13&0&0&0&0&5&0&2&4&1&1&0&0\\
Russia&ECE&7&1&0&0&0&0&0&6&0&0&0&0&0\\
Serbia&ECE&3&0&0&0&0&0&0&1&0&2&0&0&0\\
Slovenia&ECE&2&0&0&0&0&2&0&0&0&0&0&0&0\\
Ukraine&ECE&1&1&0&0&0&0&0&0&0&0&0&0&0\\
Albania&SEE&2&0&0&0&1&0&0&0&1&0&0&0&0\\
Bosnia&SEE&7&0&0&0&0&3&0&4&0&0&0&0&0\\
Cyprus&SEE&4&0&0&0&4&0&0&0&0&0&0&0&0\\
Greece&SEE&5&0&0&0&2&0&0&0&3&0&0&0&0\\
Kosovo&SEE&1&0&0&0&0&0&0&0&1&0&0&0&0\\
Macedonia&SEE&3&0&0&0&0&0&1&0&2&0&0&0&0\\
Turkey&SEE&6&0&0&0&2&0&0&0&3&0&0&0&0\\
Yugoslavia&SEE&17&0&0&0&1&6&0&2&6&0&2&0&0\\ \hline
Total&&2302&988&526&260&139&123&110&49&26&27&25&22&7\\

\hline
\end{tabular}
\end{center}
}
\noindent Table 1. Counts of Subjects from Each Country Classified to
Each Cluster.  Labels in column two create country groupings where
necessary due to small counts of subjects in many individual
countries. Country groupings NWE, ECE, and SEE include countries from north
west, east central, and south east Europe, respectively.   Eight 
clusters (A-H) were given
descriptive cluster labels based on the majority country or
country grouping membership:  (A) Swiss,
(B) British Isles,  (C) Iberian Peninsula, (D) Italian A, (E) Central, (F)
Italian B, (G) North East, and (H) South East. The remaining 4 clusters are labeled I, J, K and L.

\newpage

\begin{center}
\begin{tabular}{llllllll}
  \hline
cluster name &  \multicolumn{1}{c}{P(cluster)} & \multicolumn{5}{c}{P(case $|$ cluster)} \\
  \hline
& & \multicolumn{1}{c}{Scenario 1}& \multicolumn{1}{c}{Scenario 2}
& \multicolumn{1}{c}{Scenario 3}
& \multicolumn{1}{c}{Scenario 4}
& \multicolumn{1}{c}{Scenario 5}\\
  \hline
African-American & 0.13 & 0.33 & 0.48 & 0.26 & 0.25 & 0.33 \\
Asian Indian & 0.13 & 0.67 & 0.49 & 0.27 & 0.2 & 0.67 \\
Mexican & 0.03 & 0.2 & 0.51 & 0.27 & 0.34 & 0.2 \\
  Asian & 0.02 & 0.8 & 0.51 & 0.26 & 0.51 & 0.8 \\
Swiss   & 0.3 & 0.5 & 0.6 & 0.65 & 0.62 & 0.6 \\
British Isles & 0.17 & 0.5 & 0.8 & 0.9 & 0.9 & 0.85 \\
Iberian Peninsula & 0.07 & 0.51 & 0.05 & 0.2 & 0.45 & 0 \\
Italian A & 0.05 & 0.5 & 0.05 & 0.15 & 0.1 & 0 \\
Central European & 0.04 & 0.5 & 0.39 & 0.39 & 0.39 & 0.2 \\
Italian B & 0.04 & 0.51 & 0.21 & 0.6 & 0.6 & 0.41 \\
North East European & 0.01 & 0.5 & 0.21 & 0.46 & 0.39 & 0.21 \\
South East European & 0.01 & 0.62 & 0.29 & 0.24 & 0.19 & 0 \\
   \hline
\end{tabular}
\end{center}
\noindent Table 2.

\newpage

\begin{center}
\begin{tabular}{lrrr}
  \hline
 & 0.05 & 0.01 & 0.005 \\
  \hline
Scenario 1&&&\\
No Correction & 0.1708 & 0.0701 & 0.0477 \\
  Smartpca & 0.0494 & 0.0095 & 0.0049 \\ 
  SpectralR & 0.0522 & 0.0104 & 0.0052 \\ 
  SpectralGEM & 0.0486 & 0.0091 & 0.0044 \\
   CMH &  0.0441 & 0.0083 & 0.0041  \\
\hline
Scenario 2&&&\\
   No Correction & 0.0774 & 0.0198 & 0.0112 \\ 
  Smartpca& 0.0524 & 0.0102 & 0.0051  \\ 
  SpectralR & 0.0519 & 0.0102 & 0.0050  \\ 
  SpectralGEM &  0.0505 & 0.0096 & 0.0047  \\ 
  CMH & 0.0446 & 0.0087 & 0.0042  \\ 
\hline
Scenario 3&&&\\
  No Correction  & 0.4305 & 0.2949 & 0.2507  \\ 
  Smartpca& 0.0514 & 0.0103 & 0.0049  \\ 
  SpectralR& 0.0511 & 0.0097 & 0.0051  \\ 
  SpectralGEM & 0.0491 & 0.0096 & 0.0046  \\ 
  CMH& 0.0438 & 0.0084 & 0.0040 \\  
\hline
Scenario 4&&&\\
  No Correction  & 0.4353 & 0.2998 & 0.2564  \\ 
  Smartpca  & 0.0517 & 0.0104 & 0.0051  \\ 
  SpectralR & 0.0507 & 0.0101 & 0.0052  \\ 
  SpectralGEM & 0.0497 & 0.0097 & 0.0049  \\ 
  CMH& 0.0444 & 0.0086 & 0.0044  \\ 
\hline
Scenario 5&&&\\
  No Correction & 0.2170 & 0.1015 & 0.0734  \\ 
  Smartpca & 0.0528 & 0.0107 & 0.0053  \\  
  SpectralR & 0.0524 & 0.0103 & 0.0051  \\ 
  SpectralGEM & 0.0502 & 0.0096 & 0.0046  \\  
  CMH& 0.0434 & 0.0084 & 0.0042  \\ 
   \hline
 \end{tabular}
\end{center}
\noindent Table 3. Type I error.

\newpage

\begin{center}
\begin{tabular}{rrrrrr}
  \hline
 & Scenario 1 & Scenario 2 & Scenario 3 & Scenario 4 & Scenario 5 \\
  \hline
No Correction & 0.817 & 0.832 & 0.769 & 0.766 & 0.815 \\
  Smartpca & 0.829 & 0.808 & 0.785 & 0.784 & 0.798 \\
  SpectralR & 0.832 & 0.804 & 0.780 & 0.782 & 0.790 \\
  SpectralGEM & 0.816 & 0.775 & 0.757 & 0.754 & 0.764 \\
  CMH & 0.818 & 0.751 & 0.741 & 0.745 & 0.717 \\
   \hline
\end{tabular}
\end{center}
\noindent Table 4. Power.

\comment{
\newpage
\begin{center}
\begin{tabular}{rrrrrrrrrrrr}
  \hline
& \multicolumn{10}{c}{PCA Approach}\\
 & 1 & 2 & 3 & 4 & 5 & 6 & 7 & 8 & 9 & 10 & Average \\
  \hline
  Scenario1 & 12 & 12 & 12 & 11 & 10 & 13 & 12 & 11 & 11 & 10 & 11.4 \\
  Scenario2 & 16 & 17 & 16 & 17 & 15 & 16 & 16 & 15 & 17 & 16 & 16.1 \\
  Scenario3 & 18 & 16 & 17 & 18 & 17 & 16 & 18 & 15 & 17 & 16 & 16.8 \\
  Scenario4 & 15 & 14 & 16 & 17 & 17 & 14 & 16 & 16 & 15 & 17 & 15.7 \\
  Scenario5 & 16 & 15 & 14 & 13 & 16 & 18 & 17 & 16 & 16 & 16 & 15.7\\   
\hline
& \multicolumn{10}{c}{Spectral Graph Approach}\\
Reps & 1 & 2 & 3 & 4 & 5 & 6 & 7 & 8 & 9 & 10 & Average \\
  \hline
  Scenario1 & 7 & 7 & 7 & 7 & 7 & 7 & 7 & 7 & 7 & 7 & 7.0 \\
  Scenario2 & 7 & 7 & 6 & 7 & 7 & 6 & 6 & 7 & 7 & 6 & 6.6 \\
  Scenario3 & 7 & 6 & 7 & 6 & 6 & 7 & 7 & 6 & 7 & 6 & 6.5 \\
  Scenario4 & 7 & 6 & 7 & 7 & 6 & 7 & 6 & 6 & 6 & 6 & 6.4 \\
  Scenario5 & 7 & 6 & 7 & 7 & 7 & 7 & 6 & 7 & 7 & 7 & 6.8 \\
   \hline
\end{tabular}
\end{center}
\noindent Table Number of significant dimensions from PCA and Spectral graph approaches.
}
\newpage

\newpage

\newpage

\newpage
\newpage

\newpage

\newpage

\end{document}